# ASPECTS OF SCALING IN GRAVITATIONAL CLUSTERING

Bhuvnesh Jain

*Max-Planck-Institut für Astrophysik, 85740 Garching, Germany.*

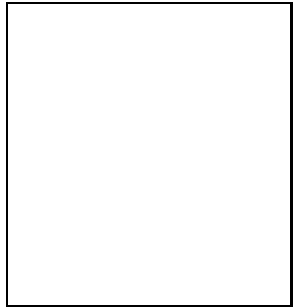

## Abstract

The scaling ansatz of Hamilton et al. effectively extends the idea of self-similar scaling to initial power spectra of any generic shape. Applications of this ansatz have provided a semi-empirical analytical description of gravitational clustering which is extremely useful. This contribution examines the two theoretical ingredients that form the basis of these applications: self-similar evolution and the stable clustering hypothesis. A brief summary of work verifying self-similar scaling for scale free spectra $P(k) \propto k^n$, with $n < -1$ is given. The main results presented here examine the hypothesis that clustering is statistically stable in time on small scales, or equivalently that the mean pair velocity in physical coordinates is zero. The mean pair velocity of particles can be computed accurately from N-body simulations via the pair conservation equation, by using the evolution of the autocorrelation function $\xi(x,t)$. The results thus obtained for scale free spectra with $n = 0, -1, -2$ and for the CDM spectrum are consistent with the stable clustering prediction on the smallest resolved scales, on which the amplitude of $\xi \gtrsim 200 - 1000$ for $n = -2$ and $n = 0$, respectively.

## 1 The Universal Scaling Ansatz

The evolution of gravitational clustering in an expanding universe has resisted rigorous analytical attempts at describing the nonlinear regime. A scaling ansatz proposed by Hamilton et al. (1991) (HKLM, reference [3]) has however provided a useful analytical prescription for computing the nonlinear autocorrelation function $\xi(x,t)$, or the power spectrum $P(k,t)$ in Fourier space. This ansatz relates the linear and nonlinear average two-point correlation functions $\bar\xi_\mathrm{L}$ and $\bar\xi_\mathrm{E}$:

$$\bar\xi_\mathrm{E}(x) = F[\bar\xi_\mathrm{L}(x_0)], \ \ x_0 = [1 + \bar\xi_\mathrm{E}(x)]^{1/3} x. \tag{1}$$

Here $\bar{\xi}(x) = (3/x^3) \int_0^x y^2 \xi(y) dy$, and $F$ is a universal function assumed to be independent of the initial spectrum. The mapping of scales from $x_0$ to $x$ takes into account the fact that nonlinear density fluctuations shrink in comoving coordinates as evolution proceeds.

A remarkable aspect of this ansatz was that the same function $F$ appeared to work for all epochs and initial spectra. HKLM found that their ansatz was in good agreement with the N-body simulations of Efstathiou et al. (1988) (EFWD, ref [2]), and used these data to determine the functional form of $F$. Motivated by this work applications and extensions of the scaling ansatz have been developed by refs [8], [10], [6] and [9]. It has been found that there is in fact a dependence on the initial spectrum in equation (1), and modified formulae that take this into account are given in ref [6].

While the functional form of $F$ in equation (1) is fitted from N-body simulations, it is important to check that the dynamical basis of the scaling ansatz is valid. The primary ingredient is self-similar scaling for scale free spectra $P(k) \propto k^n$ in an Einstein-de Sitter universe. Section 2 contains a brief discussion of the self-similar scaling of spectra with $n < -1$, based on refs [4] and [5]. In Section 3 we examine the validity of the stable clustering hypothesis which describes the deeply nonlinear regime and thus fixes the asymptotic functional form of $F$. Testing for self-similarity and stable clustering is of course important in understanding gravitational dynamics, quite independent of their applications in implementing the scaling ansatz. These issues have recently been addressed by refs [1] and [9] as well.

## 2 Self-Similar Evolution

The results of EFWD established self-similar scaling for spectra with $n \geq -1$, but they lacked the resolution to probe the $n = -2$ spectrum. Here we present a direct test of the scaling of the Fourier amplitude $\Delta(\vec{k}, t)$ of the density, defined by $\hat{\delta}(\vec{k}, t) = \Delta e^{i\phi}$ where $\hat{\delta}(\vec{k}, t)$ is the Fourier transform of the overdensity in real space, $\delta(x, t)$. This is based on work done with E. Bertschinger – a detailed analysis of the scaling of the $n = -2$ spectrum, and comparison with other recent results will be presented elsewhere. We use a $p^3m$ simulation of $128^3$ particles with the Plummer force softening parameter $\epsilon = 1/2560$ of the box size. For each mode of the Fourier density we compute the deviation of $\Delta(\vec{k}, t)$ from the linear evolution $\Delta_1 \propto a(t)$, where $a(t)$ is the expansion scale factor. By averaging the absolute values of the deviations of modes with wavenumber $k$ in the range $k_c - 0.5 < k < k_c + 0.5$, we define characteristic nonlinear scales $k_c$ as follows:

$$\left\langle \frac{|\Delta(\vec{k}_c, a) - \Delta_1(\vec{k}_c, a)|}{\Delta_1(\vec{k}_c, a)} \right\rangle = \Delta_c, \tag{2}$$

where $\Delta_c$ is a dimensionless constant of order unity. Since the magnitude of the fractional departures from linear growth for each mode within a given $k$-shell is summed, this statistic probes the scaling properties more directly than comparing a statistic like the power spectrum at different times would [4].

Figure 1 shows the results for four different values of $\Delta_c$. The self-similar scaling, $k_c \propto a(t)^{-2/(3+n)} \propto a^{-2}$ is indicated by a dotted line. Also shown is the scaling $k_c \propto a^{-1}$ which would result if the dynamics were driven by the large scale bulk velocity, a formally divergent quantity for $n < -1$. Some analytical as well as numerical evidence points to the significance of this scaling. However, the results shown in Figure 1 demonstrate that for modes with wavelength less than $1/10$ the box size, $k > 10$, the standard self-similar scaling is valid. It also provides an estimate of the effect of large-scale power for the $n = -2$ spectrum, since the growth of the amplitude of modes with wavelengths $< 1/10$ the box-size is slowed down by the absence of power on scales larger than the box.

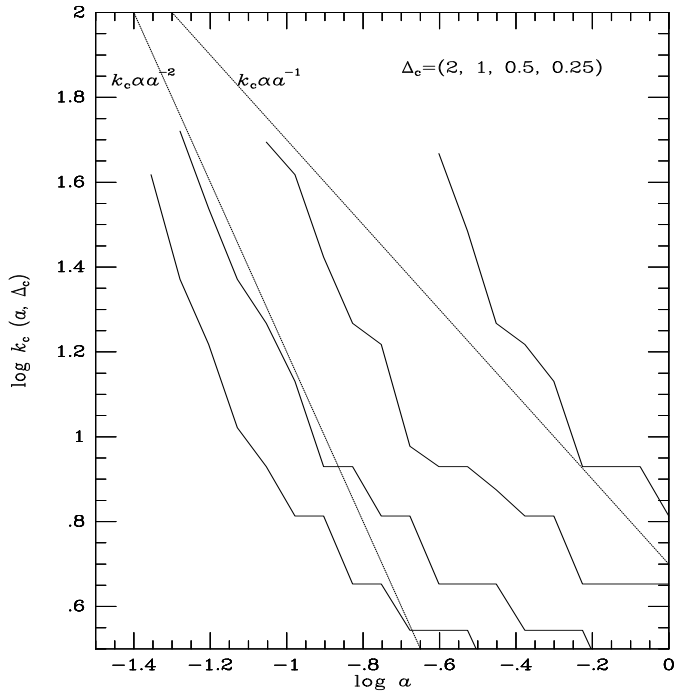

Figure 1: Scaling of the amplitude $\Delta(\vec{k},t)$ for $n=-2$. The scaling in time of characteristic wavenumber scales, derived from the departure of the amplitude from linear growth, $k_c(a,\Delta_c)$ vs. $a$ is shown for 4 different values of $\Delta_c$ (see equation (2) for the definition of $k_c(a,\Delta_c)$). The values of $a$ are in arbitrary units, while those of $k$ are in units of the box-size such that $k=1$ is the fundamental mode. All the curves have a slope close to the standard self-similar scaling, $k_c \propto a^{-2}$, shown as the dotted line on the left. For $k$ below about 10 the slope becomes shallower, due to the limitation of a finite box.

A detailed analytical analysis also reveals that while a class of statistical measures related to the phase $\phi$ would show the bulk velocity scaling, the density amplitude should scale self-similarly. Thus the conclusion of our analytical analysis in ref [5], and of a numerical study in preparation, is that self-similar scaling holds for scale free spectra with $-3 < n < 1$ (see also refs [1] and [9]).

## 3  Stable Clustering

The stable clustering hypothesis fixes the growth of the autocorrelation function $\xi(x,t)$ in the deeply nonlinear regime. This hypothesis states that on sufficiently small scales the mean relative velocity of particle pairs is zero or, equivalently, that the mean number of neighbors remains constant in time at a given physical separation. It relies on the physical picture of a virialized cluster which is no longer expanding or contracting in physical coordinates. The discussion below follows Peebles (1980) [11].

In comoving coordinates the mean pair velocity $v(x,t) = a\dot{x}$. By symmetry this velocity is directed along the line joining the pair of particles. If stable clustering is valid then the peculiar velocity must cancel the Hubble velocity on that scale, so that the net velocity in

physical coordinates is zero:
$$v(x,t) = a\dot{x} = -\dot{a}x. \tag{3}$$

To establish the connection of the above statement of the stable clustering ansatz to the growth of the correlation function, consider the equation of pair conservation, obtained by integrating the 2nd BBGKY equation over momenta (Section 71 of [11]):

$$\frac{\partial \xi}{\partial t} + \frac{1}{x^2 a}\frac{\partial}{\partial x}\left[x^2(1+\xi)v\right] = 0. \tag{4}$$

We can re-write equation (4) by integrating over the volume of a sphere, and re-arranging terms to obtain

$$\frac{a}{3(1+\xi)}\frac{\partial \bar{\xi}}{\partial a} = \frac{-v}{\dot{a}x} = \frac{-v}{Hr}, \tag{5}$$

where $Hr$ is the Hubble velocity on the physical scale $r = ax$, and $\bar{\xi}(x,t)$ is the mean interior correlation function. We shall use this form of the pair conservation equation to measure $v$ from the N-body simulations.

On substituting the assumption of equation (3) in (4), the functional form of $\xi(x,t)$ which satisfies (4) becomes restricted to

$$\xi(x,t) = a^3 f[ax], \tag{6}$$

where we have approximated $1 + \xi \simeq \xi$, as the stable clustering hypothesis is applied on very small scales were $\xi$ is at least of order 100. Equation (6) fixes the growth of $\xi$ in physical coordinates. If the initial spectrum is scale free and the universe is Einstein-de Sitter, then the functional form of equation (6) can be combined with the requirement of self-similarity to obtain the shape of the correlation function as well. The result is

$$\xi(x,t) \propto (x/a^\alpha)^{-\gamma} \; ; \; \gamma = (9+3n)/(5+n) \, , \; \alpha = 2/(3+n). \tag{7}$$

## 3.1 Testing Stable Clustering

N-body simulations provide a useful means of testing the stable clustering ansatz. The results of EFWD suggested that stable clustering might hold for $\xi \gtrsim 100$, but their simulations lacked the small scale resolution required to see evidence of it. Two simple tests of stable clustering are provided by equations (3) and (6) (and for scale free spectra, the more detailed form of equation (7)). Theoretically the two tests are equivalent, but in practice it is far easier to measure the growth and shape of $\xi(x,t)$ than it is to directly measure $v(x,t)$ as $x \to 0$. This is because of the large fluctuations in the mean pair velocity that arise due to the high velocity dispersion of a finite number of particles sampled at a given $x$.

Therefore we "measure" the mean pair velocity by solving the pair conservation equation in the form given by equation (5). Figure 2 shows $-v/Hr$ plotted as a function of $\bar{\xi}$ for four different spectra. The results are from $p^3m$ simulations with $100^3$ particles and a Plummer force softening $\epsilon = 1/2500L$, where $L$ is the box-size, for $n = 0, -1$; $128^3$ particles with $\epsilon = L/2560$ for $n = -2$; and $144^3$ particles with $L = 100$ Mpc, $h = 0.5$, and $\epsilon = 65$ kpc for the CDM spectrum. Three different output times, chosen to probe the full range of $\bar{\xi}$ accessible in the simulations, are shown for each spectrum. For CDM these are $a = 0.3, 0.5, 0.8$ with a $\sigma_8 = 1$ normalization. The results at the high-$\bar{\xi}$ end are plotted up to $x = 2\epsilon$. The near coincidence of the results for the three different times in each of the scale free spectra establishes consistency with self-similarity. It also shows how small the scatter in the results is, in contrast to a direct measurement of $v$ which would have a scatter about 5 times larger in the nonlinear regime.

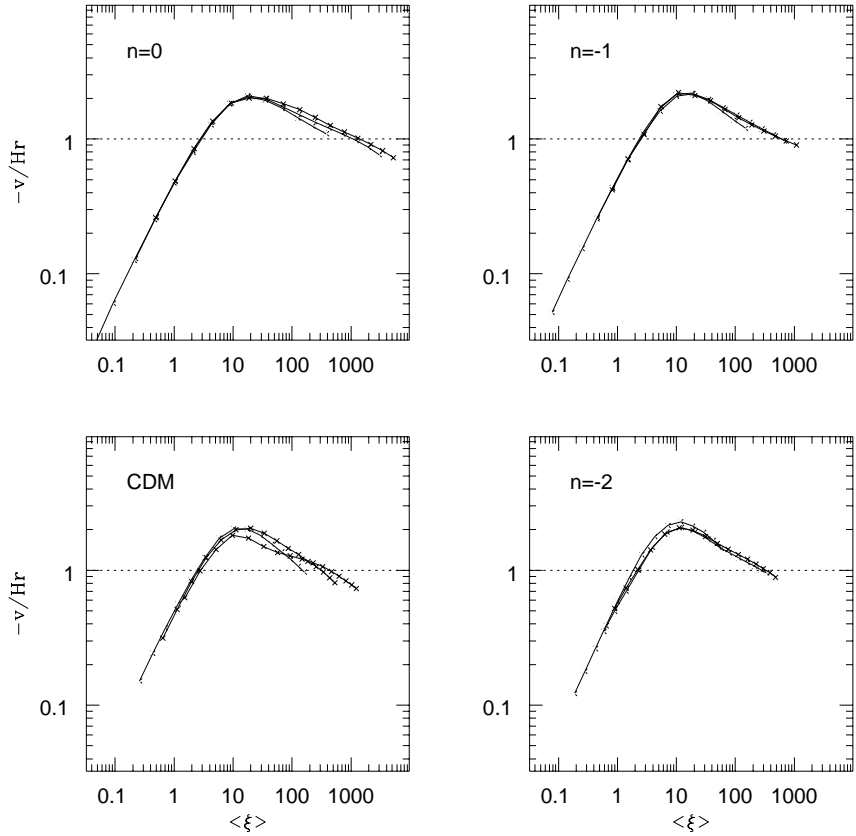

Figure 2: The ratio of the mean pair velocity to the Hubble velocity on the same scale $-v(x,t)/Hr$ vs. mean interior correlation function $\bar{\xi}(x,t)$, denoted here by $<\xi>$, at three different output times for each of the four spectra indicated in the panels. $v$ is computed by using the time evolution of $\bar{\xi}$ to solve the pair conservation equation as described in Section 3. Note that self-similarity requires that the three curves for each of the scale free spectra coincide, stable clustering that they asymptote to the value 1 at the high-$\bar{\xi}$ end, and the scaling ansatz predicts that their functional form is almost independent of the spectrum.

It is evident that at the small scale, highly nonlinear end $-v/Hr \simeq 1$. However there is no evidence for the onset of an asymptotic regime, in which $-v/Hr = 1$ for a wide enough range of scales. Larger simulations which measure small scale nonlinearity well beyond $\xi \sim 1000$ are therefore required to test whether the ratio $-v/Hr$ remains constant as $x \to 0$. Figure 2 hints that this ratio continues to decrease below the value unity, but since the last two points plotted could underestimate $-v/Hr$ due to limited numerical resolution this issue must be addressed with care, as discussed in Section 4.

## 4  Discussion

Measuring the asymptotic value of $\xi(x,t)$ or $v(x,t)$ as $x \to 0$ requires a careful analysis of the finite numerical resolution of the simulations (see ref [1] for a detailed discussion). We have computed $v(x,t)$ by using the evolution of $\xi$ rather than by measuring it directly, as the high velocity dispersion causes a direct measurement to have much larger scatter. The dominant source of error in $\xi$ at small scales is force softening which suppress its growth, and can therefore cause the inferred value of $-v/Hr$ to be artificially low. Hence the decreasing trend of $-v/Hr$ with increasing $\bar\xi$ at the smallest scales probed (between $2-4\epsilon$) should be viewed with caution. On these scales it is barely above the maximum error expected from the limited numerical resolution over the range of $\sim 4$ in $\bar\xi$ on the high-$\bar\xi$ end. Figure 2 also shows that the conclusion $-v/Hr \simeq 1$ is accurate to within $20-30\%$ for this range of scales. These results are supported by a detailed analysis of numerical resolution effects, by checking them with an ensemble of simulations for the $n = -1$ spectrum, and by other tests of stable clustering [7]; a comparison with the results of refs [1] and [9] will also made.

If one adopts the pragmatic point of view that the verification of stable clustering is useful only to the extent that it fixes the nonlinear growth of $\xi$ on a certain range of scales, then Figure 2 indicates that for $\bar\xi \gtrsim 1000$ for $n \simeq 0$, and for $\bar\xi \gtrsim 200$ for $n \simeq -2$ and CDM, the results of the simulations analyzed here are consistent with the stable clustering prediction. This is adequate for most applications such as fixing the strongly nonlinear form of $\xi$ in the formulae obtained from the scaling ansatz. However some interesting dynamical questions need to be addressed. Why is the slope of $\xi$ steeper than the stable clustering prediction, or equivalently, why is there a net infall, $-v/Hr > 1$, for $\bar\xi \lesssim 200-1000$ (for $n = -2$ and $n = 0$ respectively)? Can the spectral dependence of the results be physically explained? Does the consistency with stable clustering claimed here hold as $x \to 0$, i.e. is it indeed the asymptotic regime? While analytical models can provide an understanding of the first two questions, the existence of an asymptotic regime can only be verified by higher resolution simulations that probe the nonlinear regime well beyond $\bar\xi \sim 1000$.

**Acknowledgements.** The work presented here on self-similar scaling was done in collaboration with Ed Bertschinger. I am grateful to him and to Simon White for the use of their simulations and for useful discussions.